\documentclass[12pt,preprint]{aastex}

\shorttitle{Exoplanets}
\shortauthors{Burkert & Ida}

\begin{document}

\title{The Separation/Period Gap in the Distribution of Extrasolar Planets around Stars
with Masses $M \geq 1.2 M_{\odot}$}

\author{Andreas Burkert\altaffilmark{1} and Shigeru Ida\altaffilmark{2}}

\altaffiltext{1}{University Observatory Munich, Scheinerstrasse 1, D-81679 Munich, 
Germany}
\altaffiltext{2}{Tokyo Institute of Technology, Ookayama, Meguro-ku, Tokyo 152-8551, Japan}

\email{burkert@usm.uni-muenchen.de, ida@geo.titech.ac.jp}

\newcommand\msun{\rm M_{\odot}}
\newcommand\lsun{\rm L_{\odot}}
\newcommand\msunyr{\rm M_{\odot}\,yr^{-1}}
\newcommand\be{\begin{equation}}
\newcommand\en{\end{equation}}
\newcommand\cm{\rm cm}
\newcommand\kms{\rm{\, km \, s^{-1}}}
\newcommand\K{\rm K}
\newcommand\etal{{\rm et al}.\ }
\newcommand\sd{\partial}

\begin{abstract}

The evidence for a shortage of exosolar planets with semimajor axes $-1.1 \leq \log (a/AU) \leq -0.2$ is
investigated. It is shown that this valley results from a gap in the radial distribution
of planets, orbiting stars with masses $M_* \geq 1.2 M_{\odot}$ (the high-mass sample, HMS).
No underabundance is found for planets orbiting stars with smaller masses. The observational data
also indicate that within the HMS population it is preferentially the more massive planets with
$M \sin(i) \geq 0.8 M_J$ that are missing. Monte-Carlo simulations of planet formation
and migration are presented that reproduce the observed shortage of planets in the observed radius regime.
A dependence on the disk depletion timescale $\tau_{\rm dep}$ is found. 
The gap is more pronounced
for $\tau_{\rm dep} = 10^6 - 10^7$ yrs than for
$\tau_{\rm dep} = 3 \times 10^6 - 3 \times 10^7$ yrs. 
This might explain the observed
trend with stellar mass 
if disks around stars with masses $M_* \geq 1.2 M_{\odot}$
have shorter depletion timescales 
than those around less massive stars. 
Possible reasons for such a dependence are a decrease
of disk size and an increase of stellar EUV flux with stellar mass.

\end{abstract}

\keywords{planetary systems: formation  -- extrasolar planets}

\section{Introduction}
Within the past decade, nearly 200 extrasolar planets have been detected.
Their orbital properties and masses and the relationships between 
planetary properties and those
of the central stars are providing interesting insight into the complexity 
of planetary system formation and
offer important constraints for theoretical models.
Recently, extrasolar planets have been found around low-mass stars (M dwarfs) by
high-precision radial-velocity (Butler et al.~2004, Bonfils et al.~2005)
and microlensing surveys (Beaulieu et al.~2006, Gould et al.~2006) with indications
that Jupiter-mass planets are rare around M dwarfs (Laughlin et al.~2004, Ida \& Lin 2005).

Theoretical models have been developed to explain the statistical properties of exoplanets.
Ida \& Lin (2005) presented Monte-Carlo simulations that predict mass and semimajor axis distributions 
of extrasolar planets around stars with various masses.
Their prescription of planet formation (Ida \& Lin 2004a) is based on 
the core-accretion scenario in which it is assumed that 
Jupiter-mass gas-giant planets
form through 1) grain condensation into kilometer-sized planetesimals, 
2) runaway planetesimal coagulation (Wetherill \& Stewart 1989; 
Kokubo \& Ida 1996), 
3) oligarchic growth of protoplanetary embryos (Kokubo \& Ida 1998, 2000), 
and finally 4) gas accretion onto solid cores (Mizuno 1980; 
Bodenheimer \& Pollack 1986; Pollack et al.~1996; Ikoma et al.~2000). The models
show that the characteristic {\it mass} distribution of 
planets around
M dwarfs should be quite different from that around FGK dwarfs, 
which agrees well with the results of
microlensing surveys (Beaulieu et al.~2006; Gould et al.~2006). The core accretion model 
also predicts a strong dependence of
the likelihood for finding planets on the stellar metallicity 
(Ida \& Lin 2004b; Kornet et al.~2005) which is in
agreement with observations (Santos et al.~2000; Fischer \& Valenti 2005). 

Like the mass distribution, the {\it semimajor axis} distribution of exoplanets 
needs to be explained by any consistent theoretical model of planet formation.
One of the most puzzling questions that arose already 
when the first Jovian planets around
main sequence stars were discovered (Mayor \& Queloz 1995) is 
their wide range of orbital radii, with distances
ranging from $2 \times 10^{-2}$ AU to the observational limit of $\sim 5$ AU. 
According to the core-accretion scenario, preferred formation locations of 
Jovian planets are the outer regions of protoplanetary disks 
where the temperatures are low
enough for ice to survive (e.g. Ida \& Lin 2004a).
However, hot Jupiters have been
observed very close to the central star ($\la 0.1$AU). 
It is unlikely that these objects formed in these inner regions,
although Jovian planets can form inside the ice boundary
in relatively massive disks (Ida \& Lin 2004a).
Orbital migration due to gravitational
interaction between the planet and the gaseous disk 
(Goldreich \& Tremaine 1980;
Lin \& Papaloizou 1986a,b; Lin et al.~1996) is considered 
as the most promising mechanism to generate hot Jupiters. It starts
with relatively fast type I migration (e.g. Ward 1986; Tanaka et al. 2002).
As soon as the planet has become massive enough to open a gap in the disk it
enters the phase of type II migration where the infall rate
is connected with the viscous evolution of the disk.
Numerical simulations have explored the complex migration process
in great details (see e.g. de Val-Borro et al.~2006 for a recent comparison of
different numerical simulations). 

The Monte Carlo models of Ida \& Lin (2004b) predict
that the (differential) distribution for extrasolar
planets with radial velocities larger than 10 m/s is an increasing
linear function in log $a$ from $a \sim 0.1$ AU 
up to $\sim 1$ AU and decreases with $a$ for $a \ga 1$AU,
independent of disk metallicity, 
if the disk depletion timescale is comparable
to the disk diffusion timescale at the radius where
Jovian planets are formed (The diffusion timescale 
determines the migration speed). Very similar results were found by
Armitage et al.~(2002) and Trilling et al.~(2002). Some planets might
survive in the outer regions.
Near the inner disk edge ($\la 0.1$AU), those planets that migrate inwards
from the outer regions may pile up (Lin et al.~1996).
As a result, a gap (valley) would be created at intermediate semimajor axes 
(equivalently, intermediate periods) in the distribution.
Udry et al.~(2003) indeed point out a shortage of planets 
in the 10-100 day period range in the observed data (see also Zucker \& Mazeh 2002)
which they interpret as a signature for a transition region 
between two categories of planets
that suffered different migration scenarios.

In this paper we analyse again the valley in the period and 
separation distribution of extrasolar planets
taking into account the newest set of observational data. 
We point out that in the observational data,
the valley is much more pronounced around
F dwarfs than GK dwarfs which may reflect a dependence on
disk size and/or on the EUV flux from the central star.
Section 2 presents the observational evidence for a gap
around F dwarfs. Section 3 discusses a series of Monte-Carlo simulations that lead
to a dependence of the planet semimajor axis distribution on stellar mass.
A discussion of the results follows in section 4.

\section{A gap in semiaxis distribution  for planets around higher-mass stars}

We use the sample of extrasolar planets around normal stars, compiled by
the Extrasolar Planets Encyclopaedia 
(http://exoplanet.eu/) and 
include all planets (total number: 175) with known semimajor axes $a$ and
stellar masses $M_*$ that have been detected by radial velocity measurements.
The sample is separated into 8 bins, equally spaced
in $\log{a}$ between $-1.7 \leq \log{(a/AU)} \leq 0.7$.

We now split the sample into planets (38) orbiting stars with masses
$M_* \geq 1.2 M_{\odot}$ (which we denote the high-mass sample, HMS)
and the rest (137) which orbit less massive stars (the so called low-mass sample, LMS). 
The right panel of figure 1 shows the distribution of semimajor axes of the LMS.
The planets are homogeneously distributed for $-1.7 \leq \log{(a/AU)} < -0.2$ with a larger
frequency in the three bins with large semimajor axes, corresponding to $\log{(a/AU)} \geq -0.2$.
No clear valley is evident for $\log{(a/AU)} < -0.2$.
The distribution looks however very different for the HMS (left panel of figure 1)
where a clear lack of planets is found in the radius regime 
of $-1.1 \leq \log{(a/AU)} \leq -0.2$.

We tested the assumption that the planets in the 5 inner logarithmic
bins with $a \leq$ 0.6 AU are consistent with a homogeneous distribution.
The chi-square test of the LMS planets
leads to a likelihood of 78\% that this sample is homogeneously distributed
in $\log a$. In contrast, the chi-square statistic of the HMS planets
gives a likelihood of only 2\% that the gap in the
semimajor axis distribution is a result of statistical fluctuations.
Although the statistical significance is still 
marginal due to the low number statistics, there is a good chance that the
lack of planets for stars with
masses above 1.2 $M_{\odot}$ in the semimajor axis regime 
0.1 AU $\leq a \leq$ 0.65 AU is real, while the distribution is
much more uniform in $\log a$ for less massive stars.  

The lower two panels of figure 1 show an even more detailed analyses of 
the planets in the HMS.  In the left panel all planets with masses 
$M \sin{i} > 0.8 M_J$ are plotted.  A large gap is visible. In contrast, 
the distribution of planets with $M \sin{i} \leq 0.8 M_J$ 
is homogeneously distributed
with no gap (lower right panel of figure 1). 
Note also in the lower right panel a
lack of planets with large semimajor axes log ($a$/AU) 
$\geq -0.2$, which was also discussed by Udry et al.~(2003).

\section{Monte-Carlo simulation}

One of the possibilities to account for the different 
planetary semimajor axis distributions 
of the HMS and LMS is different sizes of their protoplanetary disks.
Armitage et al.~(2002), Trilling et al.~(2002) and
Ida \& Lin (2004b) find that the semimajor axis distributions are
controlled by the ratio of the disk depletion timescale ($\tau_{\rm dep}$)  
to the viscous diffusion timescale $\tau_{\nu}(r_1)$, evaluated at
the formation site of the Jovian planets ($r_1 \sim$ 1--10AU).
Because $\tau_{\nu}(r_1)$ determines the type II migration speed
of Jovian planets while $\tau_{\rm dep}$ determines
the duration of type II migration, 
Jovian planets migrate more and the semimajor axis
distribution is smoothed out more for larger $\tau_{\rm dep}/\tau_{\nu}(r_1)$.
For smaller $\tau_{\rm dep}/\tau_{\nu}(r_1)$, on the other hand,
Jovian planets tend to retain their original locations 
and the semimajor axis
distribution shows more clear gaps. 
We show below that $\tau_{\rm dep}/\tau_{\nu}(r_1)$ decreases
with stellar mass, leading to a more pronounced gap for higher-mass stars.

The formation of Jovian planets is favored in 
the regions just beyond the ice boundary (e.g., Ida \& Lin 2004a). It is therefore
reasonable to identify $r_1$ with the radius $r_{\rm ice}$
of the ice boundary (Hayashi 1981)
\begin{equation}
r_{\rm ice} \simeq 2.7 \left(\frac{L_*}{L_{\odot}}\right)^{1/2} {\rm AU} \simeq
2.7 \left(\frac{M_*}{M_{\odot}}\right)^{2} {\rm AU}, 
\end{equation}
where we assume that the stellar luminosity $L_*$ is proportional to $M_*^4$,
although the dependence may be slightly weaker 
for pre-main sequence stars.
We adopt a disk temperature $T$ that is regulated by 
stellar irradiation (Hayashi 1981), 
\begin{equation}
T = 280 \left(\frac{r}{1{\rm AU}}\right)^{-1/2} \left(\frac{M_*}{M_{\odot}}\right) {\rm K}.
\end{equation} 
Then the viscosity $\nu = \alpha c_s^2 /\Omega \propto \alpha r M_*^{1/2}$,
where $c_s \propto \sqrt{T}$ is the sound velocity, 
$\Omega \propto r^{-3/2}$ is the Keplerian frequency 
and $\alpha$ is the parameter of the alpha-viscosity model
(Shakura \& Sunyaev 1973). 
With these assumptions for $r_1$ and $\nu$,
\begin{equation}
\tau_{\nu}(r_1) \sim \frac{r_1^2}{3\nu(r_1)}
    \sim 10^5 
    \left(\frac{\alpha}{10^{-3}}\right)^{-1}
    \left(\frac{M_*}{M_{\odot}}\right)^{3/2}
    {\rm yrs}.
\label{eq:diff_time1}
\end{equation}
The estimate the disk depletion timescale we
adopt the standard similarity solution of viscously evolving disks
(e.g., Lynden-Bell \& Pringle 1974, Hartmann et al.~1998).
The disk depletion time is then
\begin{equation}
\tau_{\rm dep} \sim \tau_{\nu}(r_0) \sim \frac{r_0^2}{3\nu(r_0)}
    \sim 3 \times 10^6 
    \left(\frac{\alpha}{10^{-3}}\right)^{-1}
    \left(\frac{r_0}{100{\rm AU}}\right)
    \left(\frac{M_*}{M_{\odot}}\right)^{-1/2}
    {\rm yrs}.
\label{eq:diff_time0}
\end{equation}
where the disk size $r_0$ 
corresponds to the radius of the maximum viscous coupling where
the radial dependence of the disk gas surface density ($\Sigma_{\rm g}$) 
changes from $r^{-1}$ to $\exp (-r/r_0)$.  
For $t < \tau_{\nu}(r_0)$, $r_0$ is constant with time.

Recently, the dependence of protoplanetary disks on their central stars
is being observationally explored. 
One of the most remarkable dependences that has been found is that the mass accretion rate
onto the stars is proportional to the {\it square} of the stellar mass $M_*$
(Muzerolle et al.~2005; Mohanty et al.~2005; Natta et al. 2006).
Alexander \& Armitage (2006) point out that a decrease in disk size $r_0$ 
with increasing $M_*$ ($r_0 \propto M_*^{-1/2}$) can 
explain such a strong dependence.
In this case, $\tau_{\rm dep} \propto M_*^{-1}$ (Eq.~\ref{eq:diff_time0}).
From this and Eq.~(\ref{eq:diff_time1}),
$\tau_{\rm dep}/\tau_{\nu}(r_1) \propto M_*^{-5/2}$. 
Therefore, $\tau_{\rm dep}/\tau_{\nu}(r_1)$ may differ
by a factor 3 between disks around stars with $\sim 0.9M_{\odot}$
and those with $\sim 1.4M_{\odot}$.

Motivated by the above arguments, we have carried out
Monte-Carlo simulations with fixed $\alpha$ but with various $\tau_{\rm dep}$ to
determine the semimajor axis distribution of planets
with radial velocities larger than 5 m/s. The
details of the calculations of planetesimal accretion,
gas accretion onto cores, gap opening and migration are described in
Ida \& Lin (2004a, b, 2005).
Here, type II migration with $\alpha = 10^{-3}$ is included,
but type I migration is neglected.
The effect of type I migration will be 
presented in a separate paper (Ida \& Lin 2006).
We assume that the surface density of planetesimals is
$\Sigma_{\rm d} = f_{\rm disk} \Sigma_{\rm d,MMSN} \propto r^{-3/2}$,
where $\Sigma_{\rm d,MMSN}$ is that of the minimum-mass solar
nebula model (Hayashi 1981) and the scaling factor $f_{\rm disk}$
takes values of 0.1--10 (see discussion in Ida \& Lin 2004a).
On the other hand, the gas surface density is
$\Sigma_{\rm g} = f_{\rm disk} 
(r/1AU)^{1/2} \Sigma_{\rm g,MMSN} (1+t/\tau_{\rm dep})^{-3/2}
\propto r^{-1} (1+t/\tau_{\rm dep})^{-3/2}$, 
corresponding to the similarity solution at $r < r_0$. 

As long as FGK dwarfs are considered,
the formation mechanism of planets is almost
independent of $M_*$ and stellar (disk) metallicity
(Ida \& Lin 2004b, 2005). We therefore show 
results only for $M_* = 1 M_{\odot}$ and [Fe/H] = 0.2
to make the discussion simple.
With $M_* = 1 M_{\odot}$ and $\alpha = 10^{-3}$,
the viscous timescale in all simulations is
$\tau_{\nu}(r_1) \simeq 10^5$ yrs 
(Eq.~\ref{eq:diff_time1}).
Since the key parameter to determine the radial variation 
in the planet semimajor axis distribution
is $\tau_{\rm dep}/\tau_{\nu}(r_1)$, we did calculations with
$\tau_{\rm dep} = 10^6$--$10^7$ yrs, $3 \times 10^6$--$3 \times 10^7$ yrs, 
and $10^7$--$10^8$ yrs, assuming that
distributions of $\tau_{\rm dep}$ are log uniform in these ranges. 
Note that the results with shorter $\tau_{\rm dep}$ correspond to
higher $M_*$.
For $\alpha = 10^{-2}$, 10 times shorter $\tau_{\rm dep}$
produce similar results.

The upper panels in Fig.~2 show the theoretically predicted  
mass and semimajor axis distributions.
The left panel assumes 
$\tau_{\rm dep} = 10^6$--$10^7$ yrs, while the right one corresponds to
$\tau_{\rm dep} = 3 \times 10^6$--$3 \times 10^7$ yrs. We identify
the former and the latter results with planets forming around HMS and LMS,
respectively. The middle panels are histograms of 
the semimajor axis distributions.
Because we artificially terminate type II migration
at 0.04AU (Ida \& Lin 2004a), we obtain too much pile-up of hot Jupiters.
In reality, many of these hot Jupiters may actually either be consumed
(e.g., Sandquist et al.~1998) or tidally disrupted 
(e.g., Trilling et al.~1998; Gu et al.~2003) by their host stars.  
The gap at 0.1--1AU is pronounced in the result of
$\tau_{\rm dep} = 10^6$--$10^7$ yrs, while 
that of
$\tau_{\rm dep} = 3 \times 10^6$--$3 \times 10^7$ yrs
shows a flatter distribution.
Because the decline of type II migration is earlier
in the shorter $\tau_{\rm dep}$ case, 
initial formation locations are more frequently retained in the
results with $\tau_{\rm dep} = 10^6$--$10^7$ yrs than in
the case of $\tau_{\rm dep} = 3 \times 10^6$--$3 \times 10^7$ yrs.
In the lower panels, we split the histogram corresponding to
$\tau_{\rm dep} = 10^6$--$10^7$ yrs (middle left panel)
into two parts: $M \geq 0.8 M_J$ (lower left panel) and 
$M < 0.8 M_J$ (lower right panel).
Because type II migration is slower for more massive planets,
the gap is more pronounced for $M \geq 0.8 M_J$.
Thus, if the results with $\tau_{\rm dep} = 10^6$--$10^7$ years and
$3 \times 10^6$--$3 \times 10^7$ yrs represent HMS and LMS, respectively,
the theoretical predictions show a trend with stellar and planetary mass that is
consistent with the 
analysis of the observational data in section 2.

Although statistical fluctuations may still be large in the observations presented 
in Fig.~1, the rise of the mass distribution at $\sim 1$ AU
appears to be steeper than the theoretical prediction. This might be due to
photo-evaporation of disk gas due to EUV radiation from the central star
which is very effective at a few AU and could inhibit type II migration
for planets that formed beyond a few AU (Matsuyama et al.~2003). Note however that
the process of disk gap formation by photoevaporation depends critically on the
stellar EUV flux in the pre-main sequence phase which is poorly known. 
The X-ray flux from pre-main sequence stars, on the other hand,
is observed and the flux may be linearly proportional to
$M_*$ (Preibisch et al.~2005).
Hence, the EUV flux may also increase with $M_*$.
Although the dependence with stellar mass may not be strong enough,
photo-evaporation effects could make the radial
gradient of the planet mass distribution at $\sim 1$AU 
steeper for the HMS sample than for the LMS sample. More theoretical work would be required
to understand this process in greater details.

\section{Discussion}

We should start the discussion with a word of caution. Despite the fact that almost 200 exoplanets
are known, the statistics for more advanced correlations like the one discussed here
is still poor. In addition, for small numbers spurious correlations between different
variables can always be found. They would appear to have a high statistical significance which 
is however caused
by the biased selection of the data. At the moment we cannot rule out that the planet shortage
in the separation range of  0.08 AU $\leq a \leq 0.6$ AU, first discussed by
Udry et al. (2003), is such a case.

However, {\it if} this valley in period and separation distribution exists, we argue
that it is due to a lack of planets with masses M $\sin$(i) $>$ 0.8 M$_J$ orbiting around
stars with masses M$_* \geq$ 1.2 M$_{\odot}$ which provides interesting new insight on the
origin of massive planets orbiting close to their parent stars and the dependence on stellar
properties. We point out that
smaller size protoplanetary disks around more massive stars can account for a deeper valley
around these stars. If MRI turbulence is responsible for the disk viscosity,
$\alpha$ values for the viscosity would not significantly depend on stellar mass. 
Adopting a constant $\alpha$,
smaller disk sizes around more massive stars would
lead to shorter disk depletion timescales,
which prevents Jovian planets formed in the outer regions from
migrating considerably.  
Strong EUV radiation from massive (pre-main) stars
may also inhibit the migration across the regions at a few AU.
Another possibility is a larger radius of the ice boundary for massive stars that
may also inhibit smoothing-out the valley.   
The valley would then be most pronounced for planets around massive stars.

To confirm the existence of the HMS period valley and test the scenario presented here, 
more detections of extrasolar
planets around more massive stars (BA dwarfs) are needed.
If our model is correct, the valley in the period/separation
distribution of Jovian planets would become even more pronounced for even more massive stars.
Because several planets around KG giants, which were
BA dwarfs in their main sequence phase, have been discovered,
plenty of planets should also exist around BA dwarfs.
Spectroscopic observations are not easy for
BA dwarfs because of a smaller number of absorption lines
and rapid rotation.  However, in spite of the difficulty,
relatively massive planets should be detectable.  
Transit surveys may not have serious problems
to search for planets around BA dwarfs, however,
detection is limited to short-period planets.
Detailed observations of differences in disk sizes between 
T Tauri stars and Herbig Ae/Be stars are also needed to
test our model.

{\bf Acknowledgments:} A. Burkert acknowledges benefits from the activities of the RTN
network "PLANETS" supported by the European Commission under the agreement
No HPRN-CT-2002-0308. A. Burkert would like to thank the colleagues at the astronomy department
(Lick Observatory) of the University of California, Santa Cruz for their hospitality.

\clearpage

\begin{figure}[!ht]
\plotone{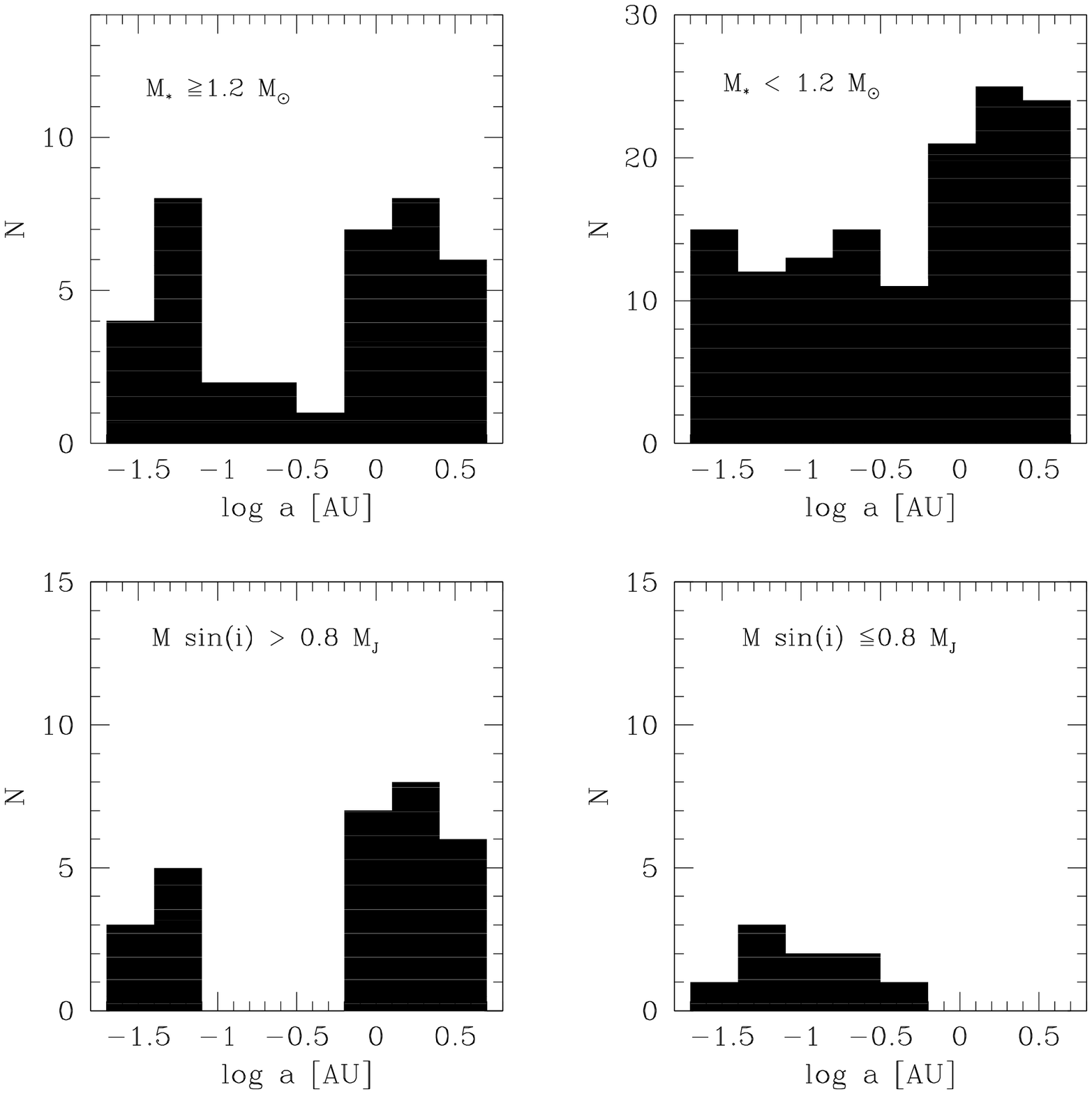}
\caption{
 \label{fig1}
The semimajor axis distribution of extrasolar planets orbiting stars
with masses $M_* \geq 1.2 M_{\odot}$ (upper left panel) and stars with masses $M_* < 1.2 M_{\odot}$
(upper right panel). The lower two panels show the distribution of extrasolar planets with masses
$M > 0.8 M_J$ (lower left panel) and $M \leq 0.8 M_J$ (lower right panel) 
for the HMS sample (upper left panel).}
\end{figure}

\begin{figure}[!ht]
\plotone{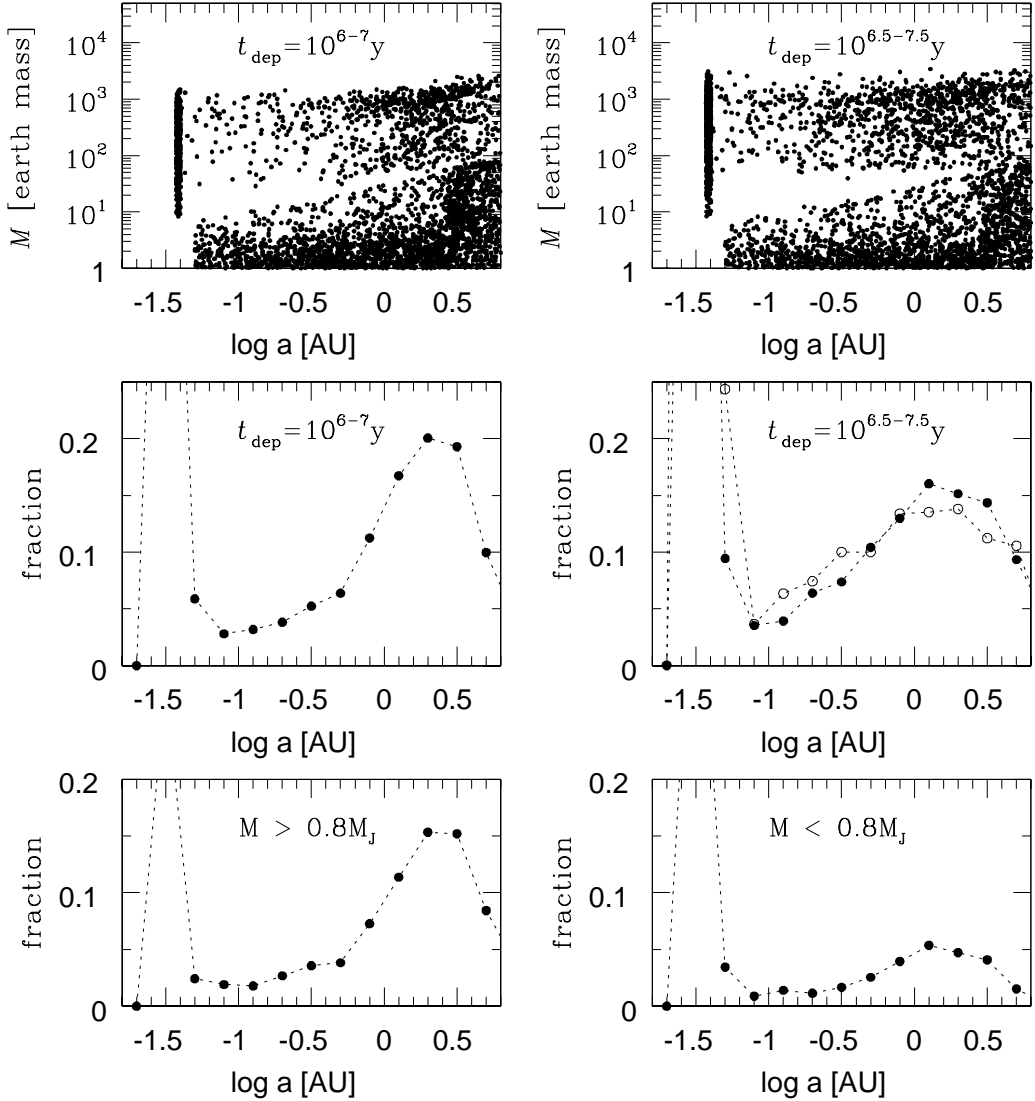}
\caption{
 \label{fig2}
The mass distributions of extrasolar planets 
predicted by the theoretical model.
Upper left and right panels show the 
mass and semimajor axis distribution with
$\tau_{\rm dep} = 10^6$--$10^7$ yrs and $3 \times 10^6$--$3 \times 10^7$ yrs,
respectively, which may correspond to HMS and LMS samples.
For other parameters, see text.
Middle panels are histograms of mass distributions of
planets with radial velocity larger than 5m/s (corresponding to
currently observable planets).
In the right panel, 
the result with $10^7$--$10^8$ yrs is
also plotted with open circles.
Lower panels are histograms
for planets with $M \geq 0.8 M_J$ (lower left panel) and 
$M < 0.8 M_J$ (lower right panel) 
for the $\tau_{\rm dep} = 10^6$--$10^7$ yrs result (middle left panel).}
\end{figure}


\begin{thebibliography}{}
\bibitem[Alexander \& Armitage (2006)]{al06}
Alexander, R.D., \& Armitage, P. 2006, ApJ, 639, L83
\bibitem[Armitage et al. (2002)]{al02}
Armitage, P., Livio, M., Lubow, S.H., \& Pringle, J.E. 2002, MNRAS, 334, 248
\bibitem[Beaulieu et al. (2006)]{be06}
Beaulieu, J.P., et al. 2006, Nature, 439, 437
\bibitem[Bodenheimer \& Pollack (1986)]{bp86}
Bodenheimer, P., \& Pollack, J.B. 1986, Icarus, 67, 391
\bibitem[Bonfils et al. (2005)]{bf05}
Bonfils, X., Forveille, R., Delfosse, X., Udry, S., Mayor, M., Perrier, C., Bouchy, F.,
Pepe, F., Queloz, D., \& Bertaux, J.L. 2005, A\&A, 443, L15
\bibitem[Butler et al. (2004)]{bv04}
Butler, P., Vogt, S., Marcy, G., Fischer, D., Wright, J., Henry, G., Laughlin, G., \& Lissauer, J.
2004, ApJ, 617, 580
\bibitem[de Val-Borro et al. (2006)]{dy06}
de Val-Borro, M., et al. 2006, MNRAS, 370, 529
\bibitem[Fischer \& Valenti (2005)]{fy05}
Fischer, F., \& Valenti, J. 2005, ApJ, 622, 1102
\bibitem[Goldreich \& Tremaine (1980)]{gt80}
Goldreich, P., \& Tremaine, S. 1980, ApJ, 241, 425
\bibitem[Gould et al. (2006)]{go06}
Gould, A., et al. 2006, ApJ, 644, L37
\bibitem[Gu et al. (2003)]{gu03}
Gu, P.G., Lin, D.N.C., \& Bodenheimer, P. 2003, ApJ, 588, 509
\bibitem[Hayashi (1981)]{ha81}
Hayashi, C. 1981, Prog. Theor. Phys. Suppl., 70, 35
\bibitem[Hartmann et al. (1998)]{ha98}
Hartmann, L., Calvet, N., Gullbring, E. \& d'Alessio, P. 1998, ApJ, 495, 385
\bibitem[Ida \& Lin (2004a)]{il04a}
Ida, S., \& Lin, D.N.C. 2004a, ApJ, 604, 388
\bibitem[Ida \& Lin (2004b)]{il04b}
Ida, S., \& Lin, D.N.C. 2004b, ApJ, 616, 567
\bibitem[Ida \& Lin (2005)]{il05}
Ida, S., \& Lin, D.N.C. 2005, ApJ, 626, 1045
\bibitem[Ida \& Lin (2006)]{il06}
Ida, S., \& Lin, D.N.C. 2006, submitted to ApJ
\bibitem[Ikoma et al. (2000)]{ik00}
Ikoma, M., Nakazawa, K., \& Emori, E. 2000, ApJ, 537, 1013
\bibitem[Kokubo \& Ida (1996)]{ki96}
Kokubo, A., \& Ida, S. 1996, Icarus, 123, 180
\bibitem[Kokubo \& Ida (1998)]{ki98}
Kokubo, A., \& Ida, S. 1998, Icarus, 131, 171
\bibitem[Kokubo \& Ida (2000)]{ki00}
Kokubo, A., \& Ida, S. 2000, Icarus, 143, 15
\bibitem[Kornet, et al. (2005)]{ko05}
Kornet, K., Bodenheimer, P., R\'ozyczka, M., \& Stepinski, T.F. 2005, A\&A, 430, 1133
\bibitem[Laughlin et al. (2004)]{la04}
Laughlin, G., Bodenheimer, P. \& Adams, F.C. 2004, ApJ, 612, L73
\bibitem[Lin \& Papaloizou (1986a)]{lp86a}
Lin, D.N.C. \& Papaloizou, J. 1986a, ApJ, 307, 395
\bibitem[Lin \& Papaloizou (1986b)]{lp86b}
Lin, D.N.C. \& Papaloizou, J. 1986b, ApJ, 309, 846
\bibitem[Lin et al. (1996)]{lb96}
Lin, D.N.C., Bodenheimer, P.  \& Richardson, D. 1996, Nature, 380, 606
\bibitem[Lynden-Bell \& Pringle (1974)]{lp74}
Lynden-Bell, D., \& Pringle, J.E. 1974, MNRAS, 168, 603
\bibitem[Matsuyama et al (2003)]{ma03}
Matsuyama, I., Johnstone, D. \& Murray, N. 2003, ApJ, 585, L143
\bibitem[Mayor \& Queloz (1995)]{mq95}
Mayor, M., \& Queloz, D. 1995, Nature, 378, 355
\bibitem[Mizuno (1980)]{mi80}
Mizuno, H. 1980, Prog. Theor. Phys. Suppl., 64, 544
\bibitem[Mohanty et al. (2005)]{mo05}
Mohanty, S., Jayawardhana, R. \& Basri, G., 2005, ApJ, 626, 498
\bibitem[Muzerolle et al. (2005)]{mu05}
Muzerolle, J., Luhman, K.L., Brice\~no, C., Hartmann, L. \& Calvet, N. 2005, 
ApJ, 625, 906
\bibitem[Natta et al. (2006)]{na06}
Natta, A., Testi, L. \& Randich, S. 2006, A\&A, 452, 245
\bibitem[Pollack et al. (1996)]{po96}
Pollack, J.B., Hubickyj, O., Bodenheimer, P., Lissauer, J.J., Podolak, M.,
\& Greenzweig, Y. 1996, Icarus, 124, 62
\bibitem[Preibisch et al. (2005)]{pr05}
Preibisch, T. et al. 2005, ApJS, 160, 582
\bibitem[Sandquist et al. (1998)]{sa98}
Sandquist, E., Taam, R.E., Lin, D.N.C. \& Burkert, A. 1998, ApJ, 506, L65
\bibitem[Santos et al. (2000)]{sa00}
Santos, N., et al. 2004, A\&A, 426, L19
\bibitem[Shakura \& Sunyaev (1973)]{ss73}
Shakura, N.I., \& Sunyaev, R.A. 1973, A\&A, 24, 337
\bibitem[Tanaka \& Ward (2002)]{tw02}
Tanaka, H., Takeuchi, T. \& Ward, W. 2002, ApJ, 565, 1257
\bibitem[Trilling et al. (1998)]{tl98}
Trilling, D.E., Benz, W., Guillot, T., Lunine, J.I., Hubbard, W.B. \& Burrows, A. 1998,
ApJ, 500, 428
\bibitem[Trilling et al. (2002)]{tl02}
Trilling, D.E., Lunine, J.I. \& Benz, W. 2002, A\&A, 394, 241
\bibitem[Udry et al. (2003)]{um03}
Udry, S., Mayor, M., \& Santos, N.C. 2003, A\&A, 407, 369
\bibitem[Ward (1986)]{wa86}
Ward, W. 1986, Icarus, 67, 164
\bibitem[Wetherill \& Stewart (1989)]{ws89}
Wetherill, G.W. \& Stewart, G.R. 1989, Icarus, 77, 330
\bibitem[Zucker \& Mazeh (2002)]{zm02}
Zucker, S. \& Mazeh, T. 2002, ApJ, 568, L113
 
\end{thebibliography}
\end{document}